\documentclass[aps,pra,amsmath,amssymb,twocolumn,superscriptaddress,notitlepage,reprint]{revtex4-1}

\pdfoutput=1
\usepackage[utf8]{inputenc}
\usepackage[T1]{fontenc}
\usepackage{graphicx}
\usepackage{dsfont}
\usepackage{color}
\definecolor{blue}{RGB}{25,51,170}
\definecolor{green}{RGB}{51, 170, 42}
\definecolor{red}{RGB}{238,25,25}
\usepackage[colorlinks,citecolor=red,linkcolor=green,urlcolor=blue]{hyperref}
\hypersetup{
  pdfkeywords={},
  pdftitle={Time-Continuous Bell Measurements},
  pdfauthor={S.G. Hofer, D.V. Vasilyev, M. Aspelmeyer, K. Hammerer}}

\input{ContinuousBellMeasurementPaper.rty}
\newcommand{\SIcite}[1]{appendix~\ref{#1}}

\begin{document}

\title{Time-Continuous Bell Measurements}
\author{Sebastian G.\ Hofer}
\email{sebastian.hofer@univie.ac.at}

\affiliation{Vienna Center for Quantum Science and Technology (VCQ), Faculty of Physics, University
  of Vienna, Boltzmanngasse 5, 1090 Vienna, Austria}
\affiliation{Institute for Theoretical Physics,
  Institute for Gravitational Physics (Albert Einstein Institute), Leibniz University Hannover,
  Callinstra\ss{}e 38, 30167 Hannover, Germany}

\author{Denis V.\ Vasilyev} \affiliation{Institute for Theoretical Physics, Institute for Gravitational
  Physics (Albert Einstein Institute), Leibniz University Hannover, Callinstra\ss{}e 38, 30167
  Hannover, Germany}

\author{Markus Aspelmeyer} \affiliation{Vienna Center for Quantum Science and Technology (VCQ),
  Faculty of Physics, University of Vienna, Boltzmanngasse 5, 1090 Vienna, Austria}

\author{Klemens Hammerer} \affiliation{Institute for Theoretical Physics, Institute for
  Gravitational Physics (Albert Einstein Institute), Leibniz University Hannover, Callinstra\ss{}e
  38, 30167 Hannover, Germany}


\begin{abstract}
  We combine the concept of Bell measurements, in which two systems are projected into a maximally entangled state, with the concept of continuous measurements, which concerns the evolution of a continuously monitored quantum system. For such time-continuous Bell measurements we derive the corresponding stochastic Schr\"odinger equations, as well as the unconditional feedback master equations.
  Our results apply to a wide range of physical systems, and are easily adapted to describe an arbitrary number of systems and measurements. Time-continuous Bell measurements therefore provide a versatile tool for the control of complex quantum systems and networks. As examples we show show that (\emph{i}) two two-level systems can be deterministically entangled via homodyne detection, tolerating photon loss up to 50\%, and (\emph{ii}) a quantum state of light can be continuously teleported to a mechanical oscillator, which works under the same conditions as are required for optomechanical ground state cooling.
\end{abstract}


\maketitle

\emph{Introduction.}---
According to the basic rules of quantum mechanics a multipartite quantum system can be prepared in an entangled state by a strong projective measurement of joint properties of its subsystems. Measurements which project a system into a maximally entangled state are called Bell measurements and lie at the heart of fundamental quantum information processing protocols, such as quantum teleportation and entanglement swapping. In systems which are amenable to strong projective measurements (\eg{}, photons \cite{bouwmeester_experimental_1997, sherson_quantum_2006} and atoms \cite{riebe_deterministic_2004, barrett_deterministic_2004}), Bell measurements constitute a well established, versatile tool for quantum control and state engineering. However, in many physical systems only weak, indirect, but time-continuous measurements are available. Over the last years a multitude of experiments have demonstrated quantum-limited time-continuous measurement and control in a range of physical systems, including single atoms \cite{bushev_feedback_2006,  kubanek_photon-by-photon_2009,  koch_feedback_2010}, cavity modes \cite{sayrin_real-time_2011, haroche_exploring_2006}, atomic ensembles \cite{smith_efficient_2006,  chaudhury_quantum_2007,  krauter_entanglement_2011}, superconducting qubits \cite{vijay_stabilizing_2012, riste_deterministic_2013}, and massive mechanical oscillators \cite{arcizet_high-sensitivity_2006,  corbitt_optical_2007,  mow-lowry_cooling_2008,  abbott_observation_2009}. Continuously monitored quantum dynamics are described through the formalism of stochastic Schr\"{o}dinger and master equations \cite{belavkin_optimal_1980, hudson_quantum_1984, gardiner_input_1985,  gardiner_wave-function_1992,  wiseman_quantum_1993,  wiseman_quantum_1994,  doherty_quantum_2000,  hopkins_feedback_2003,  jacobs_straightforward_2006,  bouten_introduction_2007,parthasarathy_introduction_1992, carmichael_open_1993, gardiner_quantum_2004, wiseman_quantum_2009,  barchielli_quantum_2009}, which in itself constitutes a cornerstone of quantum control. Surprisingly, no exhaustive connection between these important concepts---Bell measurements and time-continuous measurements---has been made so far.

\begin{figure}[t]
  \centerline{\includegraphics[width=.8\columnwidth]{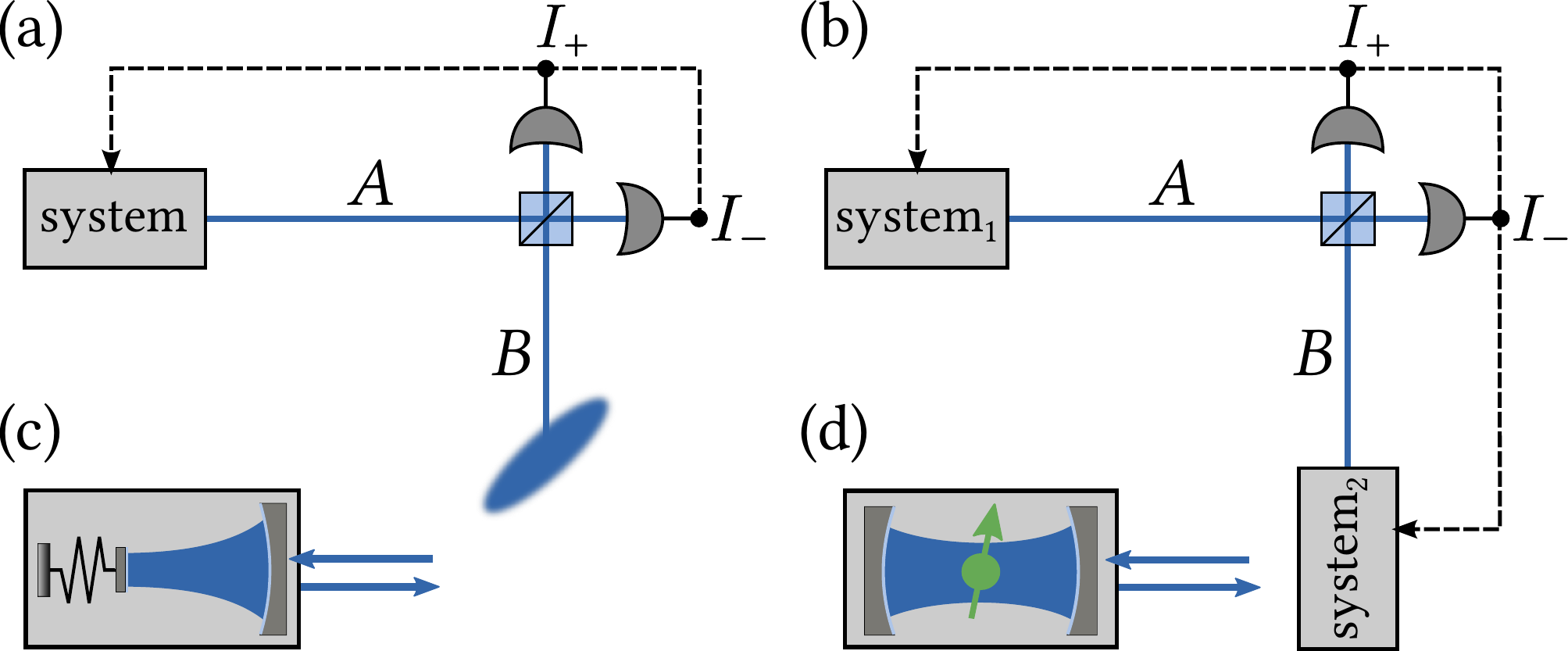}}
  \caption[]{Schematic setups for (a) time-continuous teleportation and (b) entanglement swapping. The systems may take the form of (c) a harmonic oscillator (\eg{}, an optomechanical cavity), or (d) a two-level system (\eg{}, a single spin).}
  \label{fig:setup}
\end{figure}

In this letter we establish this connection and introduce the notion of \emph{time-continuous Bell measurements} \footnote{We here use the term `Bell measurement' in the context of continuous variables, where it describes the measurement projecting onto the maximally entangled EPR states.}, which are realized via continuous homodyne detection of electromagnetic fields, and can be applied to a great number of systems, including those which cannot be measured projectively. We derive the constitutive equations of motion---the conditional stochastic Schr\"{o}dinger/master equation and the unconditional feedback master equation---of the monitored systems.
In particular we study two generic scenarios: \emph{Time-continuous quantum teleportation} of a general optical state of Gaussian (squeezed) white noise to a second system realizes a continuous remote state-preparation protocol [Fig.~\ref{fig:setup}(a)]. \emph{Continuous entanglement swapping} provides a means for dissipatively generating stationary entanglement [Fig.~\ref{fig:setup}(b)].
The corresponding fundamental equations of motion are applicable to any of the above mentioned platforms \cite{bushev_feedback_2006,  kubanek_photon-by-photon_2009,  koch_feedback_2010,sayrin_real-time_2011, haroche_exploring_2006,smith_efficient_2006,  chaudhury_quantum_2007,  krauter_entanglement_2011,arcizet_high-sensitivity_2006,  corbitt_optical_2007,  mow-lowry_cooling_2008,  abbott_observation_2009,vijay_stabilizing_2012, riste_deterministic_2013} and are the main result of this work. Along the lines of the present derivation, it is straightforward to treat different protocols, and to generalize our results to more complex setups involving an arbitrary number of systems and measurements, with applications in the continuous control of quantum networks.

To illustrate the power of our approach we demonstrate that two two-level systems can be continuously and deterministically driven to an entangled state---ideally a Bell state---through homodyne detection of light, tolerating photon losses up to 50\%. This scheme can provide the basis for a dissipative quantum repeater architecture \cite{vollbrecht_entanglement_2011}. Furthermore we show how to implement time-continuous teleportation in an optomechanical system \cite{chen_macroscopic_2013, aspelmeyer_cavity_2013} where the quantum state of continuous-wave light is continuously transferred to a moving mirror, requiring only an optomechanical cooperativity larger than one, as demonstrated in \cite{murch_observation_2008,teufel_sideband_2011,chan_laser_2011,brooks_non-classical_2012,purdy_observation_2013,safavi-naeini_squeezed_2013}.

\emph{Continuous Teleportation.}---
\label{sec-2} We consider the setup shown in Fig.~\ref{fig:setup}(a): A system $S$ couples to a 1D electromagnetic field $A$ via a linear interaction $H_{\mathrm{int}}=\ii [s\,a^{\dagger}(t)-s^{\dagger}a(t)]$, where $s$ is a system operator (\eg{} a cavity creation/destruction, or spin operator), and the light field is described (in an interaction picture at a central frequency $\omega_0$) by $a(t)=\int \dd{\omega}\, a(\omega)\, \ee^{-\ii(\omega-\omega_0)t}$. Analogously a second 1D field $B$ is described by an operator $b(t)$. Our first goal is to derive a stochastic master equation (SME) for the state of the system $S$, conditioned on the results of a time-continuous Bell measurement on the two fields $A$ and $B$. In a Markov approximation we restrict ourselves to white-noise fields, which means that both $a(t)$ and $b(t)$ are $\delta$-correlated. This allows us to introduce the \Ito{} increment $\dd{A}(t)=a(t)\dt$ (and analogously $\dd{A}^{\dagger}$, $\dd{B}$, $\dd{B}^{\dagger}$), and to express the time evolution of the state $\ket\phi$ of the overall system ($S+A+B$) as a stochastic Schrödinger equation in \Ito{} form \cite{gardiner_quantum_2004,wiseman_quantum_2009},
\begin{equation}
  \label{eq:1}
  \dd{\ket{\phi}} = \left( -\ii H_{\mathrm{eff}}\dt + s\,\dd{A^{\dagger}}\right) \ket{\phi},
\end{equation}
where $H_{\mathrm{eff}}=H_{\mathrm{sys}}-\ii \frac{1}{2}s^{\dagger}s$, with the (unspecified) system Hamiltonian $H_{\mathrm{sys}}$.

We assume that the initial state of the overall system is $\ket{\phi(0)}= \ket{\psi(0)}_{S} \ket{\mathrm{vac}}_A\ket{M}_B$, where $\ket{M}$ is an arbitrary pure Gaussian state defined by the eigenvalue equation $[(N+M^{*}+1)b(t)-(N+M)b^{\dagger}(t)]\ket{M}_B=0$. The parameters $N\in \mathbb{R}$, $M\in \mathbb{C}$ obey the relations $N\geq 0$ ($N=M=0$ true for vacuum) and $|M|^2=N(N+1)$. The white noise model essentially assumes that the squeezing bandwidth is larger than all other system time scales. Making use of the fact that $a(t)\ket{\phi(t)}=a(t)\ket{\phi(0)}=0$ and the above eigenvalue equation, we can rewrite equation \eqref{eq:1} in terms of the Einstein--Podolsky--Rosen EPR operators $X_+=(a+a^{\dagger}+b+b^{\dagger})/\sqrt{2}$ and $P_-=+\ii(a-a^{\dagger}-b+b^{\dagger})/\sqrt{2}$, which can be simultaneously measured in this setup. The resulting equation reads
\begin{equation}
  \label{eq:2}
  \dd{\ket{\phi}} =\left[ -\ii H_{\mathrm{eff}}\dt +\,s\,(\mu\, \dd{X_+} + \ii \nu\, \dd{P_-})\right]\ket{\phi},
\end{equation}
with $\mu=(1-M+M^{*})/(1+N+M^{*})$ and $\nu=\ii(1+2N+M+M^*)/(1+N+M^{*})$. Writing Eq.~\eqref{eq:1} in this form enables us to project \eqref{eq:2} onto the EPR state $\ket{I_+I_-}_{AB }$, defined by $X_+\ket{I_+I_-}_{AB }=I_{+}\ket{I_+I_-}_{AB }$ and $P_-\ket{I_+I_-}_{AB }=I_{-}\ket{I_+I_-}_{AB }$. This leads to the so-called linear stochastic Schrödinger equation \cite{carmichael_open_1993}
\begin{equation}
  \label{eq:3}
  \dd{\ket{\tilde{\psi}_{\mathrm{c}}}} =\left\{ -\ii H_{\mathrm{eff}}\dt +\,s\,[\mu I_+(t) + \ii \nu I_-(t)]\dt\right\}\ket{\tilde{\psi}_{\mathrm{c}}},
\end{equation}
for the unnormalized system state $\ket{\tilde{\psi}_{\mathrm{c}}}$, which is conditioned on the measurement results $I_{\pm}$. Note that $I_+$ and $I_-$ are real-valued, Gaussian random processes, which are proportional to the measured homodyne photocurrents. As $I_{\pm}$ result from mixing the fields $A$ and $B$ on a beam-splitter, they carry information about both fields, and will therefore be correlated, which is a crucial feature of a Bell measurement. We can write \cite{goetsch_linear_1994,wiseman_quantum_2009}
\begin{subequations}
  \label{eq:4}
  \begin{align}
    I_+(t)&=\sqrt{1/2}\mean{s+s^{\dagger}}_{\psi(t)}+\xi_{+}(t),\\
    I_-(t)&=\ii\sqrt{1/2}\mean{s-s^{\dagger}}_{\psi(t)}+\xi_{-}(t),
  \end{align}
\end{subequations}
where $\xi_{\pm}(t)=\dW_{\pm}(t)/\dt$ is zero-mean, Gaussian, white noise with corresponding Wiener increments $\dW_{\pm}$ \cite{gardiner_quantum_2004}. The (co-)variances of $\dW_\pm$ are given by
\begin{subequations}
  \label{eq:5}
  \begin{align}
    w_1\,\dt&{:=} (\dW_+)^2=[N+1+(M+M^{*})/2]\dt,\\
    w_2\,\dt&{:=} (\dW_-)^2=[N+1-(M+M^{*})/2]\dt,\\
    w_3\,\dt&{:=} \dW_+\dW_-=-[\ii(M-M^{*})/2]\dt,
  \end{align}
\end{subequations}
as follows essentially from the initial mean values with respect to the optical fields $\mean{(X_+)^2}_{\phi(0)}$, $\mean{(P_-)^2}_{\phi(0)}$, etc. As expected, we in general find non-zero cross-correlations between $I_+$ and $I_-$, which depend on the squeezing properties of the input field $B$. Using \Ito{} rules \cite{gardiner_quantum_2004} we can construct the corresponding stochastic master equation (in \Ito{} form) for the system state conditioned on the Bell measurement result,
\begin{equation}
  \label{eq:6}
  \dd{\rho_\mathrm{c}} =\mathcal{L}\rho_{\mathrm{c}}\dt+\! \frac{1}{\sqrt{2}}\!\left\{ \mathcal{H}[\mu s]\rho_{\mathrm{c}}\,\dW_{+}\!+ \mathcal{H}[\ii\nu s]\rho_{\mathrm{c}}\,\dW_{-} \right\},
\end{equation}
where we defined $\mathcal{L}\rho=-\ii[H_\mathrm{sys},\rho]+\mathcal{D}[s]\rho$, the Lindblad operator $\mathcal{D}[s]\rho=\left( s\rho s^{\dagger} - \frac{1}{2}\rho s^{\dagger}s - \frac{1}{2}s^{\dagger}s\rho \right)$, and $\mathcal{H}[s]\rho=\left( s-\mean{s} \right)\rho+\rho\left( s-\mean{s} \right)^{\dagger}$.

We now apply Hamiltonian feedback proportional to the homodyne photocurrents to the system, a scenario which covers the case of continuous quantum teleportation of the state of field $B$ to the system $S$. We follow \cite{wiseman_quantum_1993-1} in order to derive the corresponding unconditional feedback master equation. Hamiltonian feedback is described by a term \([\dot{\rho}_{\mathrm{c}}]_\mathrm{fb} = \sqrt{1/2} \left( I_+ \mathcal{K}_+ + I_{-}\mathcal{K}_-\right) \rho_\mathrm{c}\), where we define $\mathcal{K}_{\pm}\rho=-\ii[F_{\pm},\rho]$, and Hermitian operators $F_{\pm}$. After incorporating this feedback term into the SME \eqref{eq:6}, and taking the classical average over all possible measurement outcomes, $\rho=E[\rho_c]$, we arrive at the unconditional feedback master equation
\begin{multline}
  \label{eq:7}
  \dot{\rho} = -\ii \left[ H_\mathrm{sys}+(1/4)\left\{(F_{+}+\ii F_{-})s+s^{\dagger} (F_+ -\ii F_{-})\right\},\rho \right] \\ + (1/2)\big\{\mathcal{D}[s-\ii F_+]\rho +\mathcal{D}[s- F_{-}]\rho + w_3 \mathcal{D}[F_++F_{-}]\rho\\ +(w_1-w_3-1)\mathcal{D}[F_+]\rho+(w_2-w_3-1)\mathcal{D}[F_{-}]\rho\big\}.
\end{multline}
This is the main result of this section. The evolution of the system $S$ thus effectively depends on the state of the field $B$ (via $w_i$) which has never interacted with $S$, and which can in principle even change (adiabatically) in time. Eq.~\eqref{eq:7} can thus be viewed as a continuous ``remote preparation'' of quantum states.

To illustrate this point we consider the case where the target system $S$ is a bosonic mode. For a system to be amenable to continuous teleportation the system--field interaction must enable entanglement creation. We thus set $s=c^{\dagger}$, with $c$ a bosonic annihilation operator, and therefore obtain $H_{\mathrm{int}}\propto c\, a(t)+c^{\dagger}a^{\dagger}(t)$ (commonly known as two-mode-squeezing interaction). Additionally we choose $F_+=\ii(c-c^{\dagger})$ and $F_-=(c+c^{\dagger})$, which means that the photocurrents $I_+$, $I_-$ will be fed-back to the $x$ and $p$ quadrature, respectively. The resulting equation can be brought into the form
\begin{equation}
  \label{eq:8}
  \dot{\rho}=-\ii[H_{\mathrm{sys}},\rho]+(2N+1)\mathcal{D}[J]\rho,
\end{equation}
where the jump operator $J$ is determined by $J \propto -\ii(2N+1-M-M^*)x+(1+M-M^{*})p$ (with an appropriate normalization). For $H_{\mathrm{sys}}=0$ equation \eqref{eq:8} has the steady-state solution $\rho_{\mathrm{ss}}=|\psi\rangle\langle\psi|$, where $J\ket{\psi}=0$. Up to a trivial transformation this state is identical to the input state $|M\rangle\langle M|$. Note that for the vacuum case $N=M=0$ we find $J=c$, which means that, devoid of other decoherence terms, the system will be driven to its ground state. Below, we will come back to this scenario, and discuss its implementation on the basis of an optomechanical system in more detail. First, however, we consider continuous entanglement swapping [Fig.~\ref{fig:setup}(b)].

\begin{figure}[t]
  \centerline{\includegraphics[width=.8\columnwidth]{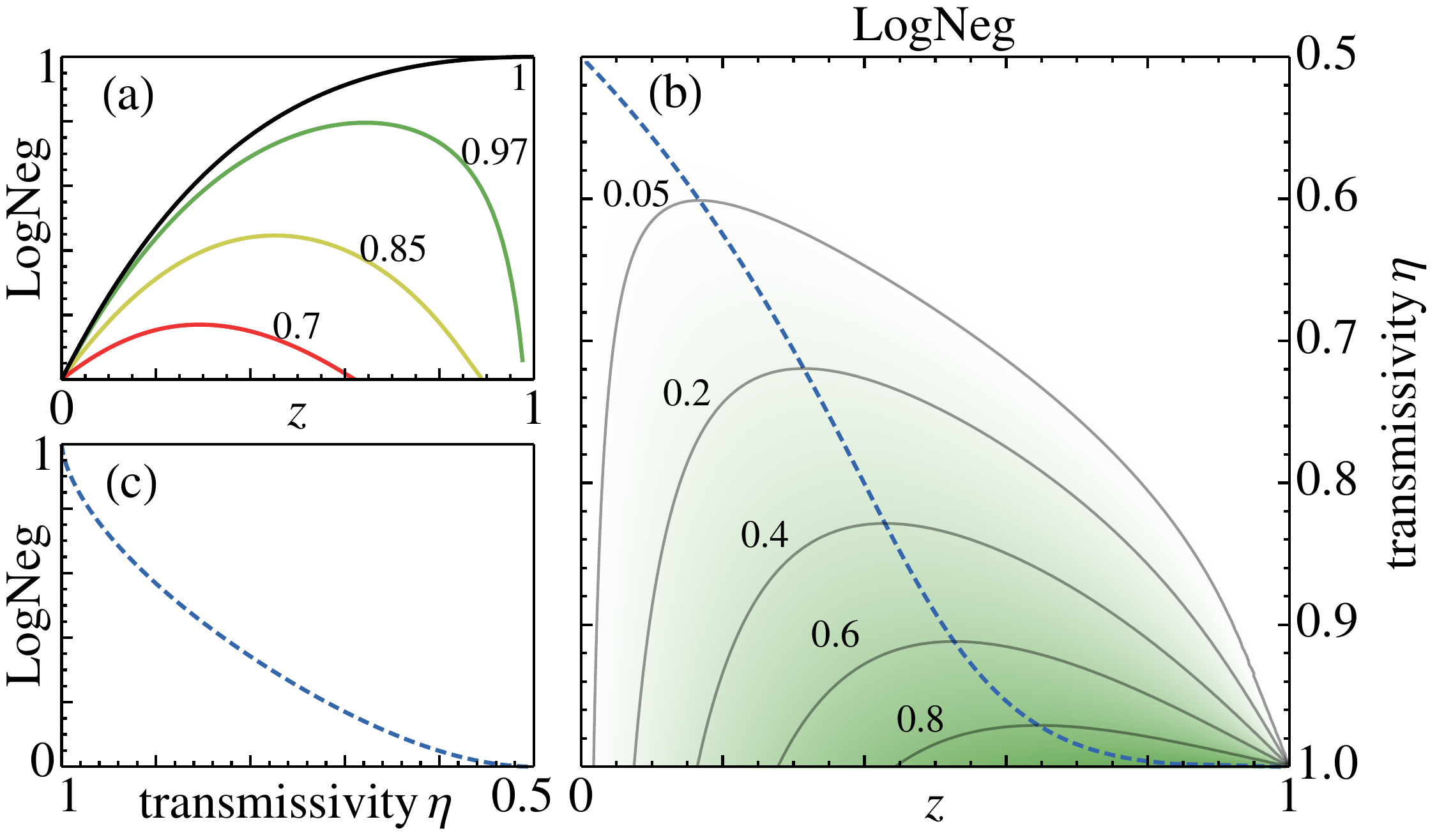}}
  \caption[]{(Color online) Performance of continuous entanglement swapping: (a) Entanglement (logarithmic negativity \cite{vidal_computable_2002}) of the stationary state versus $z$ parameterizing the light-matter interaction $\propto \ii[(\sqrt{z(1+z)}\sigma^+-\sqrt{1-z}\sigma^-)a^\dagger(t)-\mathrm{h.c.}]$, with transmissivity $\eta$ as indicated. (b) Entanglement versus $z$ and $\eta$ for optimized gains. (c) Entanglement versus transmissivity for optimized $z$. Nonzero entanglement can be achieved even for losses approaching 50\%.}
  \label{fig:entanglement}
\end{figure}

\emph{Continuous Entanglement Swapping.}---
\label{sec-3} We now replace the Gaussian input state in mode $B$ with a field state emitted by a second system, which couples to the field $B$ via $H_{\mathrm{int}}=\ii [s_2 b^{\dagger}(t)-s_2^{\dagger}b(t)]$. Using the same logic as before we can derive the linear stochastic Schrödinger equation for the bipartite state $\ket{\tilde{\psi}}$ (of $S_1$ and $S_2$)
\begin{equation}
  \label{eq:9}
  \dd{\ket{\tilde{\psi}}} =\left[ -\ii H_{\mathrm{eff}}\dt +s_+ I_+(t)\dt + \ii s_- I_-(t)\dt\right]\ket{\tilde{\psi}},
\end{equation}
where now $H_{\mathrm{eff}} = H_{\mathrm{sys}}^\mathrm{(1)}+H_{\mathrm{sys}}^\mathrm{(2)}-\frac{\ii}{2} \sum_{i=1,2}s_i^{\dagger}s_i$ and $s_{\pm}=s_1\pm s_2$. Accordingly, the homodyne currents read
\begin{subequations}
  \label{eq:10}
  \begin{align}
    I_+(t)&=\sqrt{1/2}\,\mean{s_++s_+^{\dagger}}_{\psi(t)}+\xi_{+}(t),\\
    I_-(t)&=\ii\sqrt{1/2}\,\mean{s_--s_-^{\dagger}}_{\psi(t)}+\xi_{-}(t),
  \end{align}
\end{subequations}
and the corresponding SME is
\begin{equation}
  \label{eq:11}
  \dd{\rho_\mathrm{c}} = \mathcal{L}\rho_{\mathrm{c}}\dt+\sqrt{1/2} \left\{ \mathcal{H}[s_+]\rho_{\mathrm{c}}\,\dW_{+} +\mathcal{H}[\ii s_-]\rho_{\mathrm{c}}\,\dW_{-} \right\},
\end{equation}
with $\mathcal{L}\rho = -\ii[H^\mathrm{(1)}_\mathrm{sys} + H^\mathrm{(2)}_\mathrm{sys},\rho] + \mathcal{D}[s_1]\rho + \mathcal{D}[s_2]\rho$. Here, the Wiener increments are uncorrelated and have unit variance, \ie{}, $(\dW_+)^2=(\dW_-)^2=\dt$, $\dW_+\dW_-=0$. Applying feedback to either or both of the two systems in the same way as before gives rise to
\begin{multline}
  \label{eq:12}
  \dot{\rho}=-\ii \left[ H,\rho \right]-\ii(1/4)\left[(F_{+}s_++\ii F_{-}s_-)+\mathrm{h.c.},\rho \right]\\
  + (1/2)\mathcal{D}[s_+-\ii F_+]\rho+(1/2)\mathcal{D}[s_--F_{-}]\rho.
\end{multline}
This is the desired feedback master equation for continuous entanglement swapping. For two bosonic modes with $s_i=c_i^\dagger$, applying a feedback strategy analogous to the case of teleportation above will drive the two systems to an Einstein--Podolsky--Rosen entangled stationary state. In view of Fig.~\ref{fig:setup}(b) the resulting topology comes close to a Michelson interferometer, for which a similar scheme was discussed in \cite{muller-ebhardt_entanglement_2008}. Note, however, that the central equations \eqref{eq:6}, \eqref{eq:7} and \eqref{eq:9}, \eqref{eq:12} are general and also apply to non-Gaussian systems. As a rather surprising application we will show, that a pure entangled state of two two-level systems (TLS) can be created \emph{deterministically}.

Consider two TLS which couple to 1D fields via operators $s_{1} = \sqrt{z(1+z)}\sigma_1^+ + \sqrt{1-z}\sigma_1^-$ and $s_{2} = \sqrt{z(1+z)}\sigma_2^+ - \sqrt{1-z}\sigma_2^-$ ($z\in[0,1]$). (For how to achieve this coupling see \SIcite{sec:two-level-system}.) The fields are subject to a continuous Bell measurement as depicted in Fig.~\ref{fig:setup}(b). The homodyne photocurrents $I_\pm(t)$ are used in a Hamiltonian feedback scheme to generate rotations of the TLS about their $x$ and $y$ axes according to $F_+=G_-\sigma^y_1+G_+\sigma^y_2$ and $F_-=G_+\sigma^x_1-G_-\sigma^x_2$, with gain coefficients $G_\pm=\sqrt{z/(1+z)}\left(1\pm\sqrt{z(1+z)/(1-z)}\right)$. For this choice of $s_i$ and $F_\pm$, and assuming that the levels in each TLS are degenerate (\ie{}, $H_\mathrm{sys}^{\mathrm{(}i\mathrm{)}}=0$), the jump operators in Eq.~\eqref{eq:12}, become $J_+=s_+-\ii F_+\propto j_1-\lambda j_2$ and $J_-=s_--F_-\propto j_2+\lambda j_1$, where $j_1=\sigma_1^-+z\sigma_2^+$ and $j_2=\sigma_2^-+z\sigma_1^+$, and $\lambda$ is a real coefficient. The common dark state of the jump operators $J_\pm\ket{\Phi}=j_{1,2}\ket{\Phi}=0$ is the pure entangled state $\ket{\Phi}\propto \ket{00}-z\ket{11}$ which becomes a maximally entangled Bell state for $z\rightarrow 1$ \cite{vollbrecht_entanglement_2011}. The particular linear combination of $j_{1,2}$ in $J_\pm$ is chosen such, that the state $\ket{\Phi}$ is also an eigenstate of the effective Hamiltonian $\bar{H}_\mathrm{eff}=\tfrac{1}{4}[(F_++\ii F_-)s_{1}+(F_+-\ii F_-)s_{2}+\mathrm{h.c.}]-\tfrac{\ii}{4} [J_+^\dagger J_++J_-^\dagger J_-]$ of Eq.~\eqref{eq:12}, \ie{}, $\bar{H}_\mathrm{eff}\ket{\Phi}=0$. Together, these properties guarantee that the stationary state of Eq.~\eqref{eq:12} is the pure entangled state $\ket{\Phi}$ \cite{kraus_preparation_2008}. Note that in this way entanglement is generated deterministically, in contrast to conditional schemes based on photon counting \cite{bose_proposal_1999,duan_efficient_2003,simon_robust_2003,browne_robust_2003,campbell_measurement-based_2008,vollbrecht_quantum_2009,santos_quantum_2012}. Also, it neither requires to couple nonclassical light into cavities \cite{cirac_quantum_1997,pellizzari_quantum_1997,van_enk_ideal_1997,parkins_position-momentum_2000,mancini_ponderomotive_2001,clark_unconditional_2003,kraus_discrete_2004,mancini_engineering_2004}, or a parity measurement on two qubits \cite{riste_deterministic_2013}. The necessary strong coupling of TLS to a 1D optical field can be achieved in a variety of physical systems, such as cavities \cite{raimond_manipulating_2001,miller_trapped_2005,walther_cavity_2006,haroche_exploring_2006,girvin_circuit_2009,vetsch_optical_2010,goban_demonstration_2012} or atomic ensembles \cite{hammerer_quantum_2010,saffman_quantum_2010,krauter_entanglement_2011}.

The ideal limit of Bell state entanglement ($z\rightarrow 1$ \footnote{With appropriate feedback a Bell state is also achieved in the limit $z\rightarrow 0$.}) is achieved only in the limit of infinite feedback gains $G_\pm$, as is to be expected for the present treatment. More sophisticated descriptions of feedback might relieve this restriction \cite{wiseman_quantum_2009}. However, in the relevant case including losses, the optimal feedback gains stay finite even in the present description. Assuming that all passive photon losses, such as finite transmission and detector efficiency, are combined in one transmissivity (or efficiency) parameter $\eta$, we have to apply the generalized feedback master equation from \SIcite{sec:bell-measurement-non} instead of Eq.~\eqref{eq:12}. For given $\eta$ and light matter interaction, \ie{} fixed $z$, we optimize the feedback gains $G_\pm$ in order to maximize the entanglement of the stationary state $\rho_{\mathrm{ss}}$. We keep the particular form of the feedback Hamiltonians $F_\pm$ as it preserves the Bell diagonal structure of $\rho_{\mathrm{ss}}$. Fig.~\ref{fig:entanglement} shows that entanglement can be achieved even for losses approaching 50\%, which is where the quantum capacity of the lossy bosonic channel drops to zero \cite{wolf_quantum_2007}.

\begin{figure}[ht]
  \centerline{\includegraphics[width=.8\columnwidth]{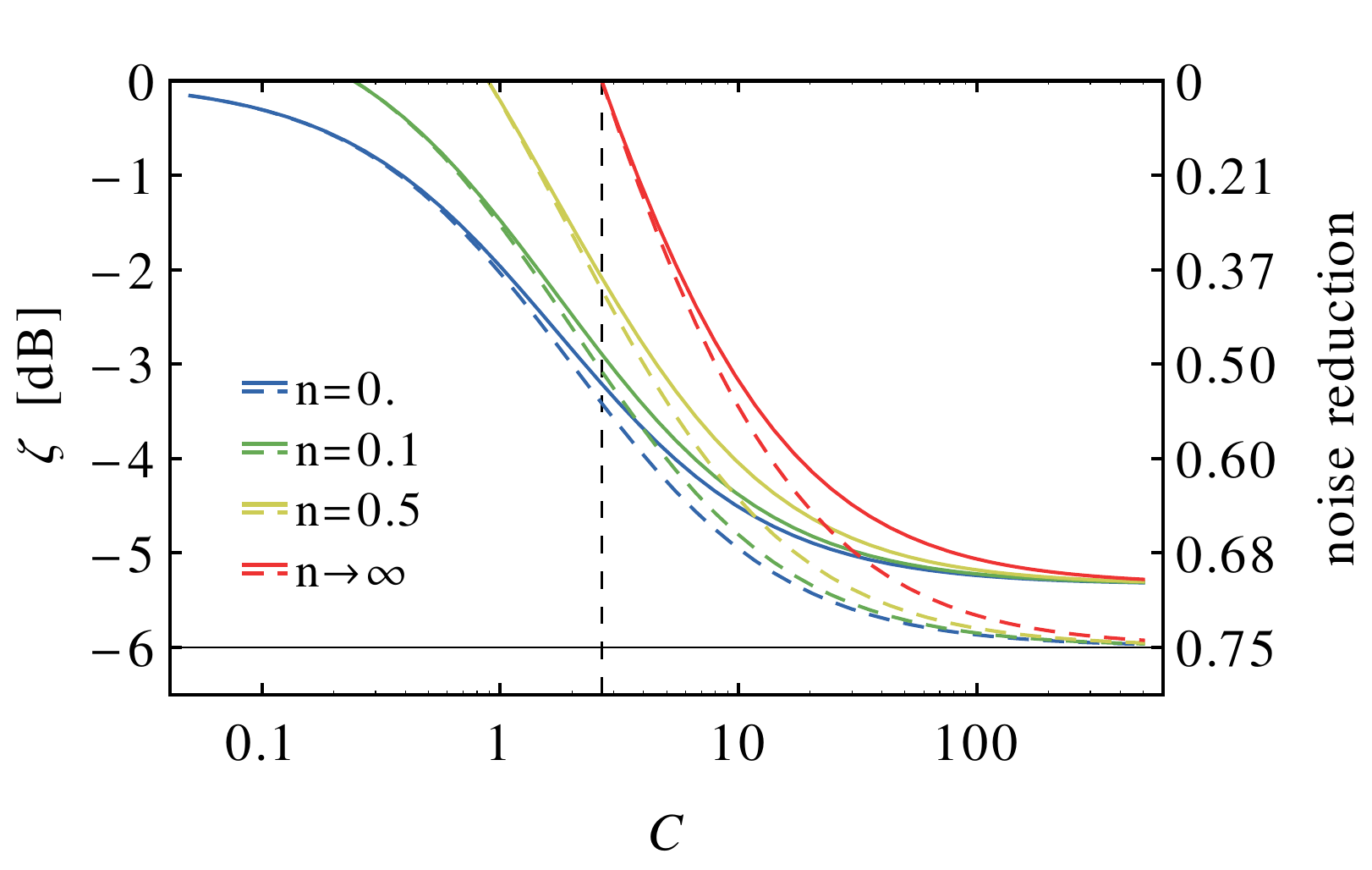}}
  \caption[]{(Color online) Mechanical squeezing $\zeta$ against cooperativity $C$ for varying mechanical bath occupation $\bar{n}=0,1/10,1/2,\infty$ (represented by different colors/gray levels) and sideband resolution $\kappa/\om=1\ (10)$ [solid (dashed) lines]. The solid black line at $\zeta=-6\, \mathrm{dB}$ shows the squeezing level of the input light (corresponding to $N\approx 0.56$). The dashed vertical line shows the value of $C_{\mathrm{crit}}=1/[\sqrt{N(N+1)}-N]\approx 2.7$ above which mechanical squeezing is achievable for any $\bar{n}$.}
  \label{fig:teleport}
\end{figure}

\emph{Application to optomechanical systems.}---
\label{sec-4} In the remainder of this article we will show how continuous quantum teleportation can be implemented in an optomechanical system in the form of a Fabry--Pérot cavity with one oscillating mirror [Fig.~\ref{fig:setup}(c)] \cite{chen_macroscopic_2013,aspelmeyer_cavity_2013}. Here the system Hamiltonian (in the laser frame at $\omega_0$) is $H_{\mathrm{sys}}=H_0+H_{\mathrm{om}}=(\omega_{\mathrm{m}}\cm^{\dagger}\cm+ \Dc\cc^{\dagger}\cc) + g(\cm+\cm^{\dagger})(\cc+\cc^{\dagger})$, where $\om$ is the mechanical frequency, $\Dc=\omega_{\mathrm{c}}-\omega_0$ is the detuning of the driving laser (at $\omega_0$) with respect to the cavity (at $\omega_{\mathrm{c}}$), and $g$ is the optomechanical coupling strength. $\cm$ and $\cc$ are bosonic annihilation operators of the mechanical and the optical mode respectively. We assume a cavity linewidth $\kappa$, a width of the mechanical resonance $\gamma$, and a mean phonon number $\bar{n}$ in thermal equilibrium.

In this system the ideal limit of continuous teleportation as given by Eq.~\eqref{eq:8} can be approached in the regime $g\ll \kappa\ll \om$ and for $\Dc=-\om$, where the resonant terms in the optomechanical interaction are $H_{\mathrm{om}}= g(\cm\cc+\cm^{\dagger}\cc^{\dagger})$. Under the weak-coupling condition ($g\ll \kappa$) the cavity follows the mechanical mode adiabatically, and we effectively obtain the required entangling interaction between the mirror and the outgoing field. The mechanical oscillator resonantly scatters photons into the lower sideband such that photons which are correlated with the mechanical motion are spectrally located at  $\omega_0-\om=\omega_{\mathrm{c}}$. Consequently, we have to modify the previous measurement setup in two ways: Firstly, we choose the center frequency of the squeezed input light at the same frequency $\omega_{\mathrm{c}}$. Secondly, we now use heterodyne detection to measure quadratures on the same sideband. These two modifications, together with the adiabatic elimination of the cavity (a perturbative expansion in $g/\kappa$ \cite{doherty_feedback_1999}) and a rotating-wave approximation (an effective coarse-graining in time \cite{gardiner_quantum_2004}), allow us to write the SME for the mechanical system, in the rotating frame at $\om$, as
\begin{multline}
  \label{eq:13}
  \dd{\rho}_\mathrm{c}=\gamma_-\mathcal{D}[\cm]{\rho}_\mathrm{c}\dt + \gamma_{+}\mathcal{D}[\cm^{\dagger}]{\rho}_\mathrm{c}\dt\\ +\sqrt{g^2\kappa/2}\left\{ \mathcal{H}[-\ii\mu\eta_+ \cm^{\dagger}]{\rho}_\mathrm{c}\dW_{+}+\mathcal{H}[\nu\eta_+ \cm^{\dagger}]{\rho}_\mathrm{c}\dW_{-} \right\},
\end{multline}
where $\eta_{\pm} = [\kappa/2+\ii(\Dc\pm\om)]^{-1}$. The first two terms describe passive cooling and heating effects via the optomechanical interaction with cooling and heating rates $\gamma_- =\gamma(\bar{n}+1)+2g^2\mathrm{Re}(\eta_-)$ and $\gamma_+ =\gamma\bar{n}+2g^2\mathrm{Re}(\eta_+)$, as was derived before in the quantum theory of optomechanical sideband cooling \cite{wilson-rae_theory_2007, marquardt_quantum_2007}. The last two terms in \eqref{eq:13} describe the continuous measurement in the sideband resolved regime for arbitrary laser detuning $\Dc$. This is an extension of the conditional master equation for optomechanical systems usually considered in the literature which concerns a resonant drive and the bad-cavity limit \cite{belavkin_measurement_1999,hopkins_feedback_2003} (see however \cite{demchenko_back_2013}).

For simplicity we assume here that we can apply feedback directly to the mechanical oscillator. We
can thus adopt the same choice of $F_{\pm}$ as before, and arrive at a feedback master equation similar to \eqref{eq:8}, $ \dot{{\rho}} = \gamma(\bar{n}+1)\mathcal{D}[\cm] {\rho} + \gamma \bar{n}\mathcal{D}[\cm^{\dagger}] {\rho} +(4g^2/\kappa) \{\lambda_{1}(\epsilon)\mathcal{D}[J_1(\epsilon)]+\lambda_2(\epsilon)\mathcal{D}[J_2(\epsilon)] \}{\rho},$ where $\epsilon=[1+(4\om/\kappa)^2]^{-1}$. (For details on how to derive $\lambda_i$ and $J_i$ refer to \SIcite{sec:diag-non-lindbl}) The protocol's performance is degraded by mechanical decoherence effects and counter-rotating terms of the optomechanical coupling, which are suppressed by $\epsilon$. For fixed input squeezing (determined by $N$) the state of the mechanical oscillator is determined by $\bar{n}$, the sideband resolution $\kappa/\om$, and the cooperativity parameter $C=g^2/(\bar{n}+1)\gamma\kappa$. In Fig.~\ref{fig:teleport} we plot the teleported mechanical squeezing $\zeta$, and compare it to the squeezing of the optical input state. As is evident from the figure there exists a critical value $C_{\mathrm{crit}}(N)=1/[\sqrt{N(N+1)}-N]$ determined by the level of input squeezing, above which mechanical squeezing can be achieved for any thermal occupation $\bar{n}$. We emphasize that this condition on the optomechanical cooperativity is essentially the same as for the recently observed ground-state cooling \cite{chan_laser_2011, teufel_sideband_2011}, back-action noise \cite{murch_observation_2008, purdy_observation_2013}, or ponderomotive squeezing \cite{brooks_non-classical_2012,safavi-naeini_squeezed_2013}. This teleportation of general Gaussian states extends previous optomechanical protocols \cite{mancini_scheme_2003, zhang_quantum-state_2003, romero-isart_optically_2011, hofer_quantum_2011} to the time-continuous domain.

\emph{Conclusion.}---
\label{sec-5} In this article we present a generalization of the standard continuous-variable Bell measurement based on homodyne detection to a continuous measurement setting. We show how this concept, together with continuous feedback, can be applied to extend existing schemes for teleportation and entanglement swapping. The presented approach can easily be extended to treat different quantum information processing protocols,  multiple measurements, and quantum networks. We suggest that the formalism developed here can serve as a basis for continuous measurement based quantum communication and information processing with both discrete and continuous variables.

\begin{acknowledgments}
  We acknowledge helpful discussions with G.\ Giedke and J.\,I.\ Cirac. We thank support provided by the European Commission (MALICIA, Q-ESSENCE, ITN cQOM), the European Research Council (ERC QOM), the Austrian Science Fund (FWF) (START, SFB FOQUS), and the Centre for Quantum Engineering and Space-Time Research (QUEST) for support. S.\,G.\ H.\ is supported by the FWF Doctoral Programme CoQuS (W1210).
\end{acknowledgments}


\appendix
\section{Bell measurement master equations}

We first treat the case of continuous teleportation. Note that due to their definition, the \Ito{} increments commute with the unitary evolution operator at all times. It thus holds that $\dd{A}(t)\ket{\phi(t)}=\dd{A}(t)\ket{\phi(0)}=\dd{A}(t)\ket{\mathrm{vac}}_{A}=0$, and by the same reasoning $[(N+M^{*}+1)\dd{B}(t)-(N+M)\dd{B}^{\dagger}(t)]\ket{\phi(t)}=0$, for the initial state $\ket{\phi(0)}=\ket{\psi(0)}_S\ket{\mathrm{vac}}_A\ket{M}_B$. Inserting these terms into \eqref{eq:1} with appropriate prefactors yields
\begin{equation*}
  \dd{\ket{\phi}} = \left\{ -\ii H_{\mathrm{eff}}\dt+s\left( \dd{A^{\dagger}}
      +\alpha\dd{A}+\dd{B}-\alpha\dd{B^{\dagger}} \right) \right\}\ket{\phi},
\end{equation*}
where $\alpha=(N+M)/(N+M^{*}+1)$. Rearranging this leads to Eq.~\eqref{eq:2}. The probability distribution of the measurement results $I_{\pm}$ is given by $\Upsilon(I_+(t),I_-(t)) = {|\langle{I_+I_-}|{\phi(t+\dt)}\rangle|}^2$ \cite{goetsch_linear_1994,wiseman_quantum_2009}, which, to first order in $\dt$, is a Gaussian with first and second moments given by \eqref{eq:4} and \eqref{eq:5} respectively.

To find the SME corresponding to \eqref{eq:3} we define $\tilde{\rho}_{\mathrm{c}}(t+\dt) = \ket{\tilde{\psi}_{\mathrm{c}}(t+\dt)}\bra{\tilde{\psi}_{\mathrm{c}}(t+\dt)}$ and note that $\ket{\tilde{\psi}_{\mathrm{c}}(t+\dt)} = \ket{\psi_{\mathrm{c}}(t)}+\dd\ket{\tilde{\psi}_{\mathrm{c}}(t)}$, where $\ket{\tilde{\psi}_{\mathrm{c}}}$ and $\tilde{\rho}_{\mathrm{c}}$ are unnormalized. After normalizing $\rho_{\mathrm{c}}(t+\dt) = \tilde{\rho}_{\mathrm{c}}(t+\dt)/\mathrm{tr}[\tilde{\rho}_{\mathrm{c}}(t+\dt)]$ we expand the resulting equation to second order in the noise increments $\dd W_{\pm}$ and apply the \Ito{} rules \eqref{eq:5} (see \cite{gardiner_quantum_2004}). With the definition $\dd\rho_{\mathrm{c}}(t)=\rho_{\mathrm{c}}(t+\dt)-\rho_{\mathrm{c}}(t)$ we find \eqref{eq:6}.

We follow the procedure developed in \cite{wiseman_quantum_1993-1} to add the feedback term $[\dot{\rho}_{\mathrm{c}}]_\mathrm{fb} = \sqrt{1/2} \left( I_+ \mathcal{K}_+ + I_{-}\mathcal{K}_-\right) \rho_\mathrm{c}$ to the conditional master equation. Note that this term must be interpreted in the Stratonovich sense \cite{wiseman_quantum_1993-1}. To reconcile it with equation~\eqref{eq:6} we thus have to convert \eqref{eq:6} to Stratonovich form, add $[\dot{\rho}_{\mathrm{c}}]_{\mathrm{fb}}$, and convert to result back to \Ito{} form. This yields
\begin{multline*}
  \dd{\rho_\mathrm{c}}= \mathcal{L}\rho_\mathrm{c}\dt+ \tfrac{1}{4}\left[ \dd{X}_{+}\mathcal{K}_{+} + \dd{P}_{-} \mathcal{K}_{-} \right]^2\rho_\mathrm{c}\\
  + \tfrac{1}{2}\left[ \dd{X}_{+}\mathcal{K}_{+} + \dd{P}_{-} \mathcal{K}_{-} \right]\left\{ \dW_{+} \mathcal{H}[\mu s]+ \dW_{-} \mathcal{H}[\ii\nu s]\right\}\rho_{\mathrm{c}}\\
  + \tfrac{1}{\sqrt{2}} \left\{ \dW_{+}\mathcal{H}[\mu s]+ \dW_{-}\mathcal{H}[\ii\nu s] \right\}\rho_{\mathrm{c}}\\
  + \tfrac{1}{\sqrt{2}} \left[ \dd{X}_{+}\mathcal{K}_{+} + \dd{P}_{-} \mathcal{K}_{-}
  \right]\rho_\mathrm{c},
\end{multline*}
where the operator ordering $\mathcal{K}\mathcal{H}$ was used in order to get a trace-preserving master equation \cite{wiseman_quantum_1993-1}. Using the fact that $\dd{X_i}\dd{X_j}=\dd{X_i}\dd{W_j}=\dd{W_i}\dd{W_j}$ together with \eqref{eq:5} and taking the average with respect to the measurement outcomes $\rho=E[\rho_{\mathrm{c}}]$ this equation can be brought into the form \eqref{eq:7}.

The case for entanglement swapping can be treated analogously. For the full system we can write
\begin{equation*}
  \begin{aligned}
    \dd{\ket{\phi}} &= \left[ -\ii H_{\mathrm{eff}}\dt+s_1 \dd{A^{\dagger}} +s_2\dd{B^{\dagger}}\right]\ket{\phi}\\
    &= \left[ -\ii H_{\mathrm{eff}}\dt+s_1\left( \dd{A^{\dagger}} + \dd{B} \right) + s_2\left(
        \dd{A} + \dd{B^{\dagger}} \right) \right]\ket{\phi},
  \end{aligned}
\end{equation*}
which, by projection onto the EPR basis, leads to \eqref{eq:9}. Note that the initial state here is assumed to be $\ket{\phi(0)}=\ket{\psi(0)}_{S_1S_2}\ket{\mathrm{vac}}_A\ket{\mathrm{vac}}_{B}$. The corresponding feedback equation is derived as for continuous teleportation, using the multiplication table $(\dd{W_+})^2=(\dd{W_-})^2=\dt$, $\dd{W_+}\dd{W_-}=0$.

\section{Bell measurement for non-unit detector efficiency}
\label{sec:bell-measurement-non}
Passive losses due to inefficient detectors or imperfect transmission can be accounted for by introducing a combined transmissivity/efficiency $0\leq \eta \leq 1$. The equations in the main text can be generalized in a straight-forward manner \cite{wiseman_quantum_2009}. In particular we find for the case of continuous teleportation the conditional master equation
\begin{equation*}
  \dd{\rho_\mathrm{c}} =\mathcal{L}\rho_{\mathrm{c}}\dt+\! \sqrt{\eta/2}\!\left\{
    \mathcal{H}[\mu s]\rho_{\mathrm{c}}\,\dW_{+}\!+ \mathcal{H}[\ii\nu s]\rho_{\mathrm{c}}\,\dW_{-}
  \right\},
\end{equation*}
and corresponding photocurrents
\begin{subequations}
  \nonumber
  \begin{align*}
    I_+(t)&=\sqrt{\eta/2}\mean{s+s^{\dagger}}_{\psi(t)}+\xi_{+}(t),\\
    I_-(t)&=\ii\sqrt{\eta/2}\mean{s-s^{\dagger}}_{\psi(t)}+\xi_{-}(t).
  \end{align*}
\end{subequations}
Including feedback as \( [\dot{\rho}_{\mathrm{c}}]_\mathrm{fb} = \sqrt{1/2\eta} \left( I_+ \mathcal{K}_+ + I_{-}\mathcal{K}_-\right) \rho_\mathrm{c}\) gives rise to the feedback master equation
\begin{multline*}
  \dot{\rho} = -\ii \left[ H_\mathrm{sys}+(1/4)\left\{(F_{+}+\ii F_{-})s+s^{\dagger} (F_+ -\ii F_{-})\right\},\rho \right]\\
  + \frac{1}{2}\bigg\{\mathcal{D}[s-\ii F_+]\rho+\mathcal{D}[s- F_{-}]+ \frac{w_3}{\eta}
  \mathcal{D}[F_++F_{-}]\\
  + \frac{w_1-w_3-\eta}{\eta}\,\mathcal{D}[F_+]+
  \frac{w_2-w_3-\eta}{\eta}\,\mathcal{D}[F_{-}]\bigg\}\rho.
\end{multline*}
Applying the same considerations to the case of entanglement swapping leads to
\begin{equation*}
  \dd{\rho_\mathrm{c}} =\mathcal{L}\rho_{\mathrm{c}}\dt+\sqrt{\eta/2} \left\{
    \mathcal{H}[s_+]\rho_{\mathrm{c}}\,\dW_{+}
    +\mathcal{H}[\ii s_-]\rho_{\mathrm{c}}\,\dW_{-} \right\},
\end{equation*}
and
\begin{multline*}
  \dot{\rho}=-\ii \left[ H,\rho \right]-\ii(1/4)\left[(F_{+}+\ii
    F_{-})s_1+\mathrm{h.c.},\rho \right] \\
  -\ii (1/4)\left[ (F_{+}-\ii F_{-})s_2+\mathrm{h.c.},\rho \right] \\
  + (1/2)\left\{\mathcal{D}[s_+-\ii F_+]\rho+\mathcal{D}[s_--F_{-}]\rho\right\}\\ + (1-\eta)/\eta
  \left\{ \mathcal{D}[F_+]\rho+ \mathcal{D}[F_-]\rho \right\},
\end{multline*}
replacing Eq.~\eqref{eq:11} and \eqref{eq:12} respectively.

\section{Diagonalization of non-Lindblad terms}
\label{sec:diag-non-lindbl}
In general the feedback master equations \eqref{eq:7} and \eqref{eq:12} are not in Lindblad form as the prefactors of the operators $\mathcal{D}$ can be negative. To cure this we can rewrite the non-unitary part of the evolution in terms of $R=(x,p)^{\mathrm{T}}$ as $\dot{\rho}=\sum_{ij}\Lambda_{ij}\left( R_i\rho R_j-\frac{1}{2} \rho R_j R_i-\frac{1}{2}R_j R_i\rho \right)$, where $\Lambda$ is a Hermitian matrix. By virtue of the eigenvalue decomposition of $\Lambda$ we can write $\dot{\rho}=\sum_i\lambda_i \mathcal{D}[J_i]\rho$ with $J_i=v_i \cdot R$, where $\lambda_i$ and $v_i$ ($i=1,2$) are the eigenvalues and eigenvectors of $\Lambda$ respectively.

\section{Two-level system interacting with light}
\label{sec:two-level-system}

\begin{figure}[t]
  \centerline{\includegraphics[width=.9\columnwidth]{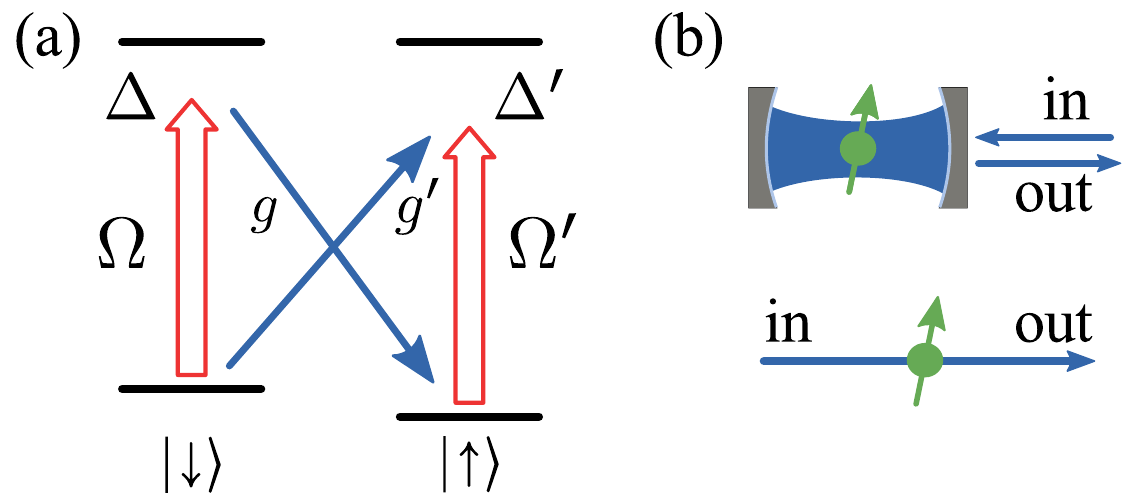}}
  \caption[]{(Color online) (a) Level scheme of a TLS interacting with a quantum field (shown by thin blue lines) with the help of driving fields $\Omega$ and $\Omega'$ (b) an adiabatic elimination of the intracavity field allows to consider the TLS directly interacting with the outside field.}
  \label{fig:levels}
\end{figure}

We explain briefly how an interaction Hamiltonian $H_{\mathrm{int}}\propto\ii [s\,a^{\dagger}(t)-s^{\dagger}a(t)]$ with $s_{1} = \sqrt{z(1+z)}\sigma_1^+ + \sqrt{1-z}\sigma_1^-$ and $s_{2} = \sqrt{z(1+z)}\sigma_2^+ - \sqrt{1-z}\sigma_2^-$ ($z\in[0,1]$), as required for continuous entanglement swapping, can be achieved. The same logic can be applied to other systems, such as a mechanical oscillator (see below). Consider an atom trapped in an optical cavity with stable ground states $|\!\downarrow\rangle$ and $|\!\uparrow\rangle$. The atom couples to the cavity through two transitions with single-photon Rabi frequencies $g$ and $g'$, and is at the same time driven by two controllable laser fields at Rabi frequencies $\Omega$ and $\Omega'$, as shown in Fig.~\ref{fig:levels}(a). If the two-photon transitions are off-resonant with detunings $\Delta\gg g,\Omega$ and $\Delta'\gg g',\Omega'$, one can eliminate the excited levels and engineer a tunable interaction of the ground states with the cavity field of the form
\begin{align*}
  H_{\mathrm{eff}} &= \ii\left(\frac{\Omega g}{\Delta}\sigma^{+}c^{\dag} + \frac{\Omega' g'}{\Delta'}\sigma^{+}c - \mathrm{h.c.}\right)\\
  &{=:}\, \ii g_\mathrm{eff}\left(sc^\dagger-s^{\dagger} c\right).
\end{align*}
Here $\sigma^{+} = |\!\!\uparrow\rangle\langle\downarrow\!\!|$ and $c$ is the cavity annihilation operator. By appropriate choice of the Rabi frequencies and detunings one can achieve any desired value of $z$ in the spin operator $s$, and at the same time set the effective coupling strength $g_\mathrm{eff}$. A tunable light--matter interaction based on two (effective) Raman transitions was recently demonstrated in \cite{krauter_entanglement_2011}. If the cavity decay (given by $\kappa$) is fast on the time scale of this effective coupling, the cavity field can be adiabatically eliminated. This gives rise to dynamics in which the spin effectively couples directly to the outside field, described by white noise operators $[a(t),a^\dagger(t')]=\delta(t-t')$ [Fig.~\ref{fig:levels}(b)]. The Hamiltonian then is of the general form
\begin{equation*}
  H_{\mathrm{eff}} =  \ii \frac{g_\mathrm{eff}}{\sqrt{\kappa}}\left[sa^\dagger(t)-s^{\dagger}a(t)\right]
\end{equation*}
considered in the main article. The procedure of adiabatic elimination is discussed in some more detail for optomechanical systems in the following section.

\section{Continuous optomechanical teleportation}
To derive Eq.~\eqref{eq:13} we start from the SME describing heterodyne detection of the optomechanical cavity's output light. It can be obtained from equation \eqref{eq:6} by replacing $s\rightarrow \cc \,\ee^{\ii \Delta_{\mathrm{lo}} t}$, where $\Delta_{\mathrm{lo}}$ is the local oscillator detuning, and adding decoherence terms for the mechanical subsystem, which are due to its coupling to a thermal bath. After going into an interaction picture with $H_0=\omega_{\mathrm{m}}\cm^{\dagger}\cm+ \Dc\cc^{\dagger}\cc$ we can adiabatically eliminate the cavity mode, by expanding the equations in the small parameter $g/\kappa$ up to first order \cite{doherty_feedback_1999, wilson-rae_cavity-assisted_2008}. Under this approximation and after setting $\Delta_{\mathrm{lo}}=\om$ the SME takes the form
\begin{multline*}
  \dd\rho_{\mathrm{c}}=\mathcal{L}_{\mathrm{c}}\rho\dt-g^2 \left[ \cm+\cm^{\dagger},
    y\rho_\mathrm{c} -\rho_{\mathrm{c}} y^{\dagger} \right]\dt + \\
  \sqrt{g^2\kappa/2}\left\{ \mathcal{H}[-\ii \mu \ee^{\ii \om t}y]\rho_{\mathrm{c}}\,\dd{W}_{+} +
    \mathcal{H}[\nu \ee^{\ii \om t}y]\rho_{\mathrm{c}}\,\dd{W}_{-} \right\}
\end{multline*}
where $\mathcal{L}_{\mathrm{m}}\rho =-\ii[\omega_{\mathrm{m}}\cm^{\dagger}\cm,\rho] + \gamma(\bar{n}+1)\mathcal{D}[\cm]\rho +\gamma\bar{n}\mathcal{D}[\cm^{\dagger}]\rho$, and $y=\eta_- \cm+\eta_+ \cm^{\dagger}$. The steady-state mean amplitude of the intracavity field is $\mean{\cc}=-\ii g\mean{y}$. The heterodyne photocurrents can thus be obtained from \eqref{eq:4} by the replacement $\mean{s}\rightarrow -\ii g\mean{y}\ee^{\ii \om t}$, thus
\begin{subequations}
  \begin{align*}
    I_+&=-\ii\sqrt{g^2\kappa/2}\mean{y\,\ee^{\ii \om t}-\mathrm{h.c.}}(t) + \xi_+(t)\\
    I_-&=\sqrt{g^2\kappa/2}\mean{y\,\ee^{\ii \om t}+\mathrm{h.c.}}(t) + \xi_-(t)
  \end{align*}
\end{subequations}
To apply the rotating wave approximation we go to the rotating frame of the mechanical oscillator (in the following denoted by $\tilde{\rho}_{\mathrm{c}}$) \footnote{Note that the mechanical resonance frequency shifts due to the optical spring effect. The rotating frame will therefore rotate at a frequency \(\omega_{\mathrm{m}}^{\mathrm{eff}}\)} and take a time average over an interval $\delta t$, which comprises many mechanical periods, but is short on the system timescales in the rotating frame, \ie{}, $1/g\gg\delta t\gg 1/\om$. Under this assumption we can pull $\tilde{\rho}_{\mathrm{c}}$ out from under the integral and drop terms rotating at a frequency $\om$. At the same time we treat $\delta t$ as infinitesimal on the system's timescale and therefore replace $\delta t\rightarrow \dt$. This leaves us with the equation
\begin{multline*}
  \dd\tilde{\rho}_\mathrm{c}=\gamma_-\mathcal{D}[\cm]\tilde{\rho}_{\mathrm{c}}\dt +
  \gamma_{+}\mathcal{D}[\cm^{\dagger}] \tilde{\rho}_{\mathrm{c}}\dt\\
  +\sqrt{g^2\kappa/2}\left\{ \mathcal{H}[-\ii\mu\eta_+
    \cm^{\dagger}]\tilde{\rho}_{\mathrm{c}}\dd{W}_{0+}+\mathcal{H}[\nu\eta_+
    \cm^{\dagger}]\tilde{\rho}_{\mathrm{c}}\dd{W}_{0-} \right\}.
\end{multline*}
We also introduced the time-averaged Wiener increments $\dW_{0\pm}=\int \dW_{\pm}$, which approximately fulfill \eqref{eq:5}, as far as $\tilde{\rho}_{\mathrm{c}}$ is concerned. In principle this equation contains additional measurement terms corresponding to sideband modes at frequencies $\pm 2\om$, which in a RWA are not correlated to the DC modes $\dW_{0\pm}$ and were therefore neglected. By renaming $\dd{W}_{0\pm}\rightarrow \dd{W}_{\pm}$ we arrive at \eqref{eq:13}. Note that by applying the RWA to the decoherence and the measurement terms consistently we assure that the resulting equation is a valid Belavkin equation \cite{belavkin_quantum_1992}.

To do feedback we do the same coarse-graining procedure for the photocurrents $I_{0\pm}=\int I_{\pm}\dt$, and find
\begin{subequations}
  \begin{align*}
    I_{0+}&=-\ii\sqrt{g^2\kappa/2}\mean{\eta_+\cm^{\dagger}-\mathrm{h.c.}}(t) + \xi_+(t),\\
    I_{0-}&=\sqrt{g^2\kappa/2}\mean{\eta_+\cm^{\dagger}+\mathrm{h.c.}}(t) + \xi_-(t).
  \end{align*}
\end{subequations}
Within these approximations the system is equivalent to the generic case in the main text. Applying the same feedback procedure we obtain
\begin{multline*}
  \dot{\tilde{\rho}}_{\mathrm{m}}=\left\{ \gamma(\bar{n}+1)\mathcal{D}[\cm] + \gamma
    \bar{n}\mathcal{D}[\cm^{\dagger}] \right\} \tilde{\rho}_{\mathrm{m}}\\
  +(4g^2/ \kappa)\left\{ (1+\epsilon) \mathcal{D}[c_{\mathrm{m}}] + w_3 \mathcal{D}[x_{\mathrm{m}}+p_{\mathrm{m}}]\right.\\
  + \left. (w_1-w_3-1)\mathcal{D}[p_{\mathrm{m}}]+(w_2-w_3-1)\mathcal{D}[x_{\mathrm{m}}]
  \right\}\tilde{\rho}_{\mathrm{m}},
\end{multline*}
In view of the previous section we can diagonalize this equation to obtain
%
\setcounter{equation}{0}
\begin{multline}
  \label{eq:26}
  \dot{\tilde{\rho}} = \left\{ \gamma(\bar{n}+1)\mathcal{D}[\cm] + \gamma
    \bar{n}\mathcal{D}[\cm^{\dagger}] \right\} \tilde{\rho}\\
  +(4g^2/\kappa)
  \left\{\lambda_{1}(\epsilon)\mathcal{D}[J_1(\epsilon)]+\lambda_2(\epsilon)\mathcal{D}[J_2(\epsilon)]
  \right\}\tilde{\rho},
\end{multline}
where $\lambda_i$ and $J_i$ are obtained from the eigenvalue decomposition of
\begin{equation*}
  \Lambda=
  \begin{pmatrix}
    w_2-\frac{1}{2}(1+\epsilon)&-w_3+\frac{\ii}{2}(1+\epsilon)\\
    -w_3-\frac{\ii}{2}(1+\epsilon)&w_1-\frac{1}{2}(1+\epsilon)
  \end{pmatrix},
\end{equation*}
where $\epsilon=[1+(4\om/\kappa)^2]^{-1}$. In the limit $\epsilon\rightarrow 0$ Eq.~\eqref{eq:26} reduces to \eqref{eq:8}, apart from the decoherence terms of the mechanical subsystem, which counteract the squeezing of the mechanical mode by driving it towards a thermal state. Operating the protocol in a regime of strong cooperativity $g^2/\kappa \gg \gamma(\bar{n}+1)$ suppresses these perturbative effects.


\bibliography{ContinuousBellMeasurement}
\end{document}